\documentclass[twocolumn,pre,aps,superscriptaddress]{revtex4-2}
\usepackage[utf8]{inputenc}
\usepackage{bm}
\usepackage[dvipsnames]{xcolor}
\usepackage{multirow}
\usepackage{amssymb}
\usepackage{amsbsy}
\usepackage{amsmath}
\usepackage{stmaryrd}
\usepackage{graphicx}
\usepackage{epsfig}
\usepackage{placeins}
\usepackage[normalem]{ulem}
\usepackage{bbold}
\usepackage{braket}
\usepackage{blindtext}

\usepackage{xcolor}
\definecolor{midnight3}{HTML}{4a6d88}

\usepackage[colorlinks,linkcolor=midnight3,citecolor=midnight3,urlcolor=midnight3]{hyperref}
\usepackage{filecontents}

\usepackage{scalerel}
\usepackage{tikz}
\usetikzlibrary{calc}
\usetikzlibrary{patterns}
\usetikzlibrary{svg.path}
\definecolor{orcidlogocol}{HTML}{A6CE39}
\tikzset{
  orcidlogo/.pic={
    \fill[orcidlogocol] svg{M256,128c0,70.7-57.3,128-128,128C57.3,256,0,198.7,0,128C0,57.3,57.3,0,128,0C198.7,0,256,57.3,256,128z};
    \fill[white] svg{M86.3,186.2H70.9V79.1h15.4v48.4V186.2z}
                 svg{M108.9,79.1h41.6c39.6,0,57,28.3,57,53.6c0,27.5-21.5,53.6-56.8,53.6h-41.8V79.1z M124.3,172.4h24.5c34.9,0,42.9-26.5,42.9-39.7c0-21.5-13.7-39.7-43.7-39.7h-23.7V172.4z}
                 svg{M88.7,56.8c0,5.5-4.5,10.1-10.1,10.1c-5.6,0-10.1-4.6-10.1-10.1c0-5.6,4.5-10.1,10.1-10.1C84.2,46.7,88.7,51.3,88.7,56.8z};
  }
}

\newcommand\orcid[1]{\href{https://orcid.org/#1}{\mbox{\scalerel*{
\begin{tikzpicture}[yscale=-1,transform shape]
\pic{orcidlogo};
\end{tikzpicture}
}{|}}}}

\makeatletter
\newcommand*{\balancecolsandclearpage}{
  \close@column@grid
  \clearpage
  \twocolumngrid
}
\makeatother

\makeatletter
\def\maketitle{
\@author@finish
\title@column\titleblock@produce
\suppressfloats[t]
\let\and\relax
\let\affiliation\@gobble@opt@one
\let\address\affiliation
\let\author\@gobble
\@author@init
\let\@authors\@empty
\let\@authors@curr\@empty
\let\@affil@list\@empty
\let\keywords\@gobble
\let\@keywords\@empty
\let\email\@gobble
\let\@address\@empty
\let\thanks\@gobble
\titlepage@sw{ %
\clearpage
}{}%
}
\makeatother

\begin{document}

\title{Real-time broadening of bath-induced density profiles
\\ from closed-system correlation functions}

\author{Tjark Heitmann \orcid{0000-0001-7728-0133}}
\email{tjark.heitmann@uos.de}
\affiliation{Department of Mathematics / Computer Science / Physics, 
University of Osnabr\"uck, D-49076 Osnabr\"uck, Germany}

\author{Jonas Richter \orcid{0000-0003-2184-5275}}
\affiliation{Department of Physics, Stanford University, 
Stanford, CA 94305, 
USA}
\affiliation{Institut f\"ur Theoretische Physik, Leibniz 
Universit\"at 
Hannover, Appelstra\ss e 2, 30167 Hannover, Germany}

\author{Jacek Herbrych \orcid{0000-0001-9860-2146}}
\affiliation{Institute of Theoretical Physics, Faculty of Fundamental Problems 
of Technology, Wroc\l aw University of Science and Technology, 50-370 Wroc\l aw, 
Poland}

\author{Jochen Gemmer}
\affiliation{Department of Mathematics / Computer Science / Physics, 
University of Osnabr\"uck, D-49076 Osnabr\"uck, Germany}

\author{Robin Steinigeweg \orcid{0000-0003-0608-0884}}
\email{rsteinig@uos.de}
\affiliation{Department of Mathematics / Computer Science / Physics, 
University of Osnabr\"uck, D-49076 Osnabr\"uck, Germany}

\date{\today}

\begin{abstract}
The Lindblad master equation is one of the main approaches to open quantum 
systems. While it has been widely applied in the context of condensed matter 
systems to study properties of steady states in the limit of long times, the 
actual route to such steady states has attracted less attention yet. Here, we 
investigate the nonequilibrium dynamics of spin chains with a local coupling to 
a single Lindblad bath and analyze the transport properties of the induced 
magnetization. Combining typicality and equilibration arguments with stochastic 
unraveling, we unveil for the case of weak driving that the dynamics in the open 
system can be constructed on the basis of correlation functions in the closed 
system, which establishes a connection between the Lindblad approach and linear 
response theory at finite times. In this way, we provide a particular example
where closed and open approaches to quantum transport agree strictly. We 
demonstrate this fact numerically for the spin-1/2 XXZ chain at the isotropic 
point and in the easy-axis regime, where superdiffusive and diffusive
scaling is observed, respectively.
\end{abstract}
\maketitle

\section{Introduction}
Understanding the dynamics of many-body quantum 
systems has seen remarkable progress in recent years \cite{Polkovnikov2011}, 
including the origin of thermalization and hydrodynamics under unitary time 
evolution \cite{Nandkishore2015, Dalessio2016, Khemani2018, Rakovszky2018}, the 
possibility of weak and strong forms of ergodicity breaking \cite{Abanin2019, 
Serbyn2021}, and the emergence of universality far from equilibrium 
\cite{Dziarmaga2010, Sieberer2013, Pruefer2018, Erne2018, Rodriguez-Nieva2022}. 
In addition to theoretical breakthroughs, these and related areas have also 
profited immensely from experiments like seminal quantum simulators, where both 
closed and open systems can be probed \cite{Kaufman2016, Luschen2017, 
Rubio-Abadal2019}. The competition of internal quantum dynamics, dissipation, 
and external driving opens up a vast landscape of exotic nonequilibrium 
phenomena~\cite{Diehl2008, Weimer2010}.

In systems with a conservation law, e.g., spin models with conserved total 
magnetization, a key role is played by the slow relaxation of the corresponding 
hydrodynamic modes \cite{Bertini2021}. While chaotic systems are typically 
expected to exhibit diffusion \cite{Lux2014, Bohrdt2017, Richter2018a}, 
anomalous transport can occur, e.g., in the presence of long-range interactions 
\cite{Kloss2019, Schuckert2020,Richter2022}, in disordered and kinetically 
constrained systems \cite{Luitz2017a, Richter2022a, Singh2021}, or in the case 
of integrable models \cite{Bulchandani2021}. For the latter, the concept of 
generalized hydrodynamics provides a powerful framework to predict the emerging 
transport behavior \cite{Bertini2016, Castro-Alvaredo2016}. In generic systems, 
in contrast, extracting quantitative values of transport coefficients like 
diffusion constants remains a formidable challenge even for sophisticated 
numerical techniques \cite{Leviatan2017, Ye2020, White2018, Rakovszky2022}.

\begin{figure}[tb]
\centering
\vspace{-.2cm}

\begin{tikzpicture}
  \node[anchor=south west,inner sep=0] (img) at (0,0) 
  {\includegraphics[width=.75\columnwidth]{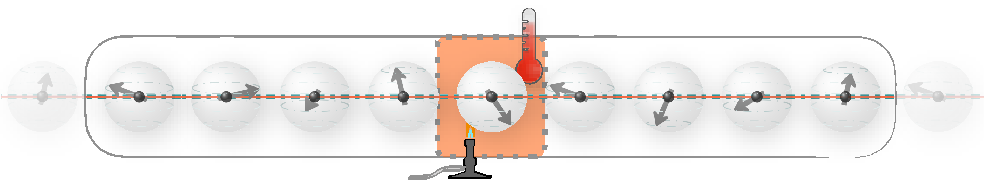}};
  \node at (-15pt,25pt) {(a)};
\end{tikzpicture}
\vspace{-.2cm}

\includegraphics[width=0.95\columnwidth]{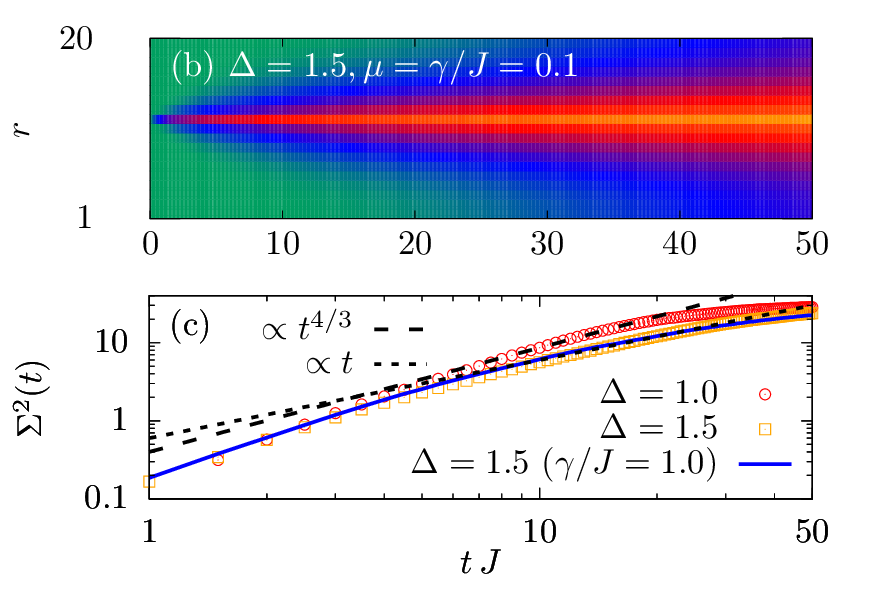}
\caption{(a) Sketch of our setup.
(b) Magnetization dynamics ${\langle S_r^z(t)\rangle}$ 
in the spin-1/2 
XXZ chain coupled to a single Lindblad bath, obtained from the full stochastic 
unraveling for anisotropy ${\Delta = 1.5}$, small coupling ${\gamma / J = 
0.1}$, weak driving ${\mu = 0.1}$, and ${N = 20}$ sites. 
(c) Corresponding 
spatial variance $\Sigma^2(t)$ for ${\Delta = 1.0}$ and ${\Delta = 1.5}$. 
Additionally, a curve for large ${\gamma/J = 1.0}$ is depicted for ${\Delta = 
1.5}$. The dashed (dotted) fits indicate superdiffusive (diffusive) scaling. 
The saturation of $\Sigma^2(t)$ at long times is due to finite $N$.}
\label{fig:variance}
\end{figure}

A canonical approach to quantum transport in closed spin or Hubbard type models 
is given by linear response theory (LRT) in the form of 
equilibrium correlation functions \cite{Bertini2021}.
A number of efficient numerical methods have been used to evaluate 
such correlation functions either in real time or in the frequency domain, 
including exact diagonalization \cite{Heidrich-Meisner2007}, matrix product 
state techniques \cite{Karrasch2012, Ljubotina2017}, Lanczos methods 
\cite{Long2003}, dynamical quantum typicality \cite{Steinigeweg2014, 
Elsayed2013, Richter2019b, Heitmann2020, Jin2021}, semiclassical approximations 
\cite{Wurtz2018}, or quantum Monte Carlo \cite{Grossjohann2010}.

An alternative approach to transport is to consider an open-system setting, 
where the model is connected at its ends to reservoirs, which 
drive a current through the bulk 
\cite{Michel2003, Wichterich2007, Znidaric2011, Znidaric2016}. The time 
evolution is often described by a Lindblad master equation which induces a 
nonequilibrium steady state at long times. State-of-the-art algorithms to 
solve the Lindblad equation are based on a matrix-product-operator formulation,
which gives access to huge system sizes, e.g., on the 
order of hundreds of spin-1/2 degrees of freedom \cite{Prosen2009, 
Verstraete2004, Zwolak2004, Weimer2021, Lenarcic2020}. Especially for systems 
in the thermodynamic limit, it is expected that the specific form and strength 
of the system-bath coupling become irrelevant for the steady state. However, 
the involved Lindblad operators in practice often have to be chosen 
heuristically. Moreover, extra care has to be taken in the case of finite 
systems to reproduce the correct behavior of the actual closed system of 
interest \cite{Prelovsek2022}. 
While agreement  
between boundary-driven transport and LRT has numerically been observed for 
selected examples \cite{Steinigeweg2009a, 
Znidaric2018}, there is no general proof that both approaches need to agree 
\cite{Kundu2009, Purkayastha2018, Purkayastha2019, Znidaric2019, Bertini2021}, 
also at weak driving.

In this Letter, we make a significant step forward to bridge the conceptual gap 
between closed-system and open-system 
numerical approaches to quantum transport. Focusing 
on the case of weak driving and relying on typicality and equilibration 
arguments, we establish a connection between LRT and the finite-time dynamics of 
an open quantum system in a simple setting introduced below 
and sketched in Fig.\ \ref{fig:variance}(a). 
Specifically, 
we unveil that open-system dynamics can be constructed from closed-system 
correlation functions. This novel connection entails both physical implications 
regarding the transport properties and consequences regarding efficient 
numerical simulations of open systems. 
We also note that Green-Kubo type relations connecting 
equilibrium correlation functions to open-system transport have been obtained 
before in classical systems, see e.g., Refs.\ \cite{Kundu2009, Dhar_2008}. We 
stress that our results and our framework are distinct from such 
approaches. Rather we provide a means to understand individual trajectories in 
the unraveling of Lindblad master equations from the dynamics of the closed 
system.

\section{Setup}
While our theoretical framework applies more generally to other 
systems, we here demonstrate its validity for the spin-1/2 XXZ chain as a 
timely example,
\begin{equation}
H = J \sum_{r=1}^N (S_r^x S_{r+1}^x + S_r^y S_{r+1}^y + \Delta S_r^z 
S_{r+1}^z) \, ,
\end{equation}
where $S_r^{x,y,z}$ are spin-1/2 operators at site $r$, ${J > 0}$ is the 
antiferromagnetic coupling constant, $\Delta$ denotes the anisotropy in 
$z$ direction, and $S_{N+1}^{x,y,z}\equiv S_1^{x,y,z}$. The high-temperature 
spin-transport properties of the integrable XXZ chain have been in the focus of 
intense theoretical and experimental efforts in recent years. While normal 
diffusion emerges for ${\Delta > 1}$ \cite{Bertini2021}, transport is 
superdiffusive at ${\Delta = 1}$ with spatiotemporal correlations following the 
Kardar-Parisi-Zhang (KPZ) scaling function (see e.g.~\cite{Bulchandani2021, 
Ljubotina2017, Ljubotina2019, Gopalakrishnan2019a}).

In this Letter, we consider a nonequilibrium situation, where the system of
interest is coupled to an external bath, as described by the Lindblad equation
\begin{equation} \label{Eq::Lind1}
\dot{\rho}(t) = {\cal L} \, \rho(t)  = i [\rho(t),H] + {\cal D} \, \rho(t) \, ,
\end{equation}
which consists of a coherent time evolution of the density matrix $\rho(t)$ and 
an incoherent damping term
\begin{equation} \label{Eq::Lind2}
{\cal D} \, \rho(t) = \sum_j \alpha_j \Big ( L_j \rho(t) L_j^\dagger - 
\frac{1}{2} \{ \rho(t), L_j^\dagger L_j \} \Big )\ , 
\end{equation}
with non-negative rates $\alpha_j$, Lindblad operators $L_j$, and the 
anticommutator $\{ \bullet, \bullet \}$. While the derivation of 
Eqs.~(\ref{Eq::Lind1}) and (\ref{Eq::Lind2}) can be a subtle task for a given 
microscopic model \cite{Wichterich2007, DeRaedt2017} (and might not always be 
justified \cite{Thingna2013, Tupkary2022}), it is the most general form of a 
time-local quantum master equation, which maps a density matrix to a density 
matrix \cite{Breuer2007}. Here, we focus on arguably the simplest possible 
setup, see Fig.\ \ref{fig:variance}(a), where $H$ is 
coupled 
to the bath at a 
single lattice site,
\begin{align}
L_1 = S_{r_0}^+ \, , \quad &\alpha_1 = \gamma (1 + \mu) \, ,
\label{Eq:1} \\
L_2 = L_1^\dagger = S_{r_0}^- \, , \quad &\alpha_2 = \gamma
(1 - \mu) \, , \label{Eq:2}
\end{align}
where $\gamma$ is the system-bath coupling, $\mu$ is the driving strength, 
and $L_1$ and $L_2$ are local Lindblad operators at site ${r_0 = N/2}$. (This 
site is arbitrary due to periodic boundaries). Note that 
throughout our work and consistent with the literature
on transport \cite{Bertini2021}, we refer to the influence of Lindblad
operators as driving. This type of
incoherent driving should not be confused with a coherent
driving by a time-dependent Hamiltonian.

Considering a homogeneous 
initial state $\rho(0)$ and choosing $\mu > 0$, excess magnetization is induced 
at the bath site and then transported through the chain. Specifically, we study 
the time evolution of local densities ${\langle S_r^z(t) \rangle = 
\text{tr}[\rho(t) S_r^z]}$, see Fig.~\ref{fig:variance}(b), which depends on 
the parameters of the system $H$, but also on the bath parameters $\gamma$ and 
$\mu$. The emerging transport behavior reflects itself in the growth of the  
spatial variance \cite{Bertini2021}
\begin{equation}
\Sigma^2(t) = \sum_r \frac{\langle S_r^z(t) \rangle}{\langle S^z(t) \rangle} 
r^2 - \Big [ \sum_r \frac{\langle S_r^z(t) \rangle}{\langle S^z(t) \rangle} r 
\Big ]^2\ , 
\end{equation}
with ${\langle S^z(t) \rangle = \sum_r \langle S_r^z(t) \rangle}$. 
Importantly, as shown in Fig.~\ref{fig:variance}(c), we find that at weak 
driving ${\mu = 0.1 \ll 1}$, the transport behavior of the isolated XXZ chain
carries over to the behavior of the open system with diffusive scaling 
(${\Sigma^2(t) \propto t}$) at ${\Delta = 1.5}$ and superdiffusive KPZ scaling 
(${\Sigma^2(t) \propto t^{4/3}}$) at ${\Delta = 1.0}$. A key contribution of the 
present work is to show how this result can be understood by connecting the 
Lindblad setting to the dynamics of correlation functions in the closed system.

\begin{figure}[tb]
\centering
\includegraphics[width=0.9\columnwidth]{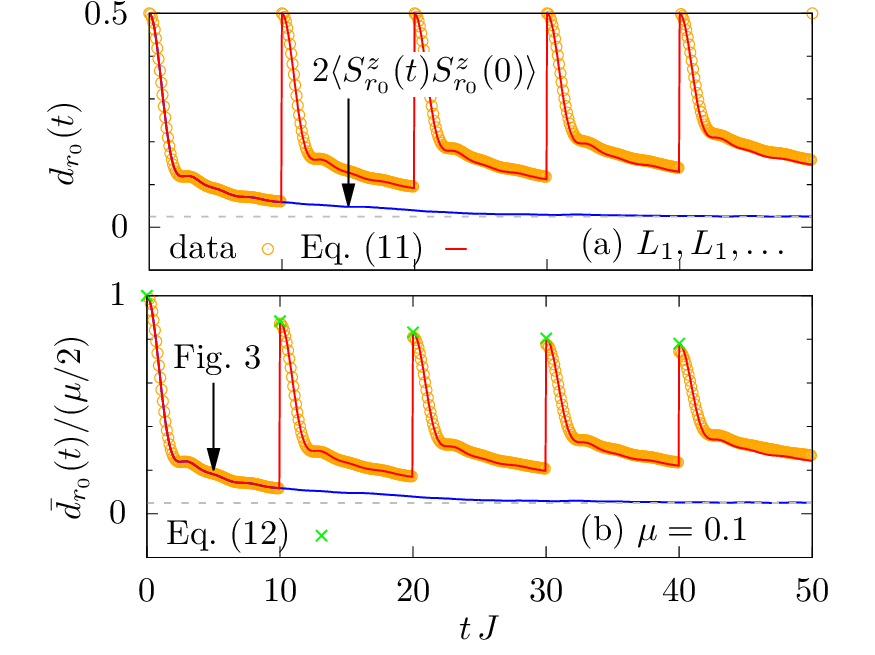}
\caption{Test setting with artificial jump times $\tau_k = 0,10,\dots$. (a) 
Magnetization dynamics $d_{r_0}(t)$ at $r_0=N/2$ for a single trajectory with 
Haar-random initial state  $| \psi(0) \rangle$ and weak driving ${\mu \ll 1}$. 
We here consider only the single Lindblad operator $L_1$. (b) Average over all 
possible trajectories with jump operators $L_1$ and $L_2$, weighted with the 
respective probabilities for ${\mu = 0.1}$. In each case, numerical data 
(circles) are found to agree convincingly with the prediction in 
Eqs.~(\ref{eq:superposition}) (curves) and (\ref{eq:amplitudes}) (crosses). 
Other parameters: ${\Delta = 1.5}$ and ${N = 20}$. The dashed line indicates 
the long-time equilibration value of the correlation function, i.e., $0.5/N$.}
\label{fig:peak}
\end{figure}

\section{Trajectories and weak Lindblad driving}
One possibility to solve the 
Lindblad equation is given by the concept of stochastic unraveling, which 
relies on pure states $\ket{\psi}$ rather than density matrices 
\cite{Dalibard1992, Michel2008}. It consists of an alternating sequence of 
stochastic jumps and deterministic evolutions with respect to an effective 
Hamiltonian ${H_\text{eff} = H  - (i/2) \sum_j \alpha_j \, L_j^\dagger L_j}$. 
Given Eqs.~(\ref{Eq:1}) and (\ref{Eq:2}), $H_\text{eff}$ here takes on the form
\begin{equation}\label{Eq::HEff}
H_\text{eff} = H  - \frac{i}{2} \gamma (1 + \mu) + i \gamma \mu n_{r_0}
\approx H  - \frac{i}{2} \gamma \, ,
\end{equation}
where ${n_{r_0} = S_{r_0}^+ S_{r_0}^- = S_{r_0}^z} + 1/2$, and the 
approximation in the last step applies for weak driving ${\mu \ll 1}$. In 
particular, for ${\mu \ll 1}$, the deterministic evolution
$\exp(-iH_\text{eff}t)\ket{\psi(0)}$ simplifies,
\begin{equation} \label{Eq::TEVO}
| \psi(t) \rangle \approx e^{- \gamma t/2} \, e^{-i H t} \, | \psi(0) \rangle 
\, ,
\end{equation}
i.e., apart from the scalar damping term, the dynamics is generated by the 
closed system $H$ only. This simplification will be central to derive our 
analytical prediction below. However, in our numerical simulations, we always 
take into account the full expression of $H_\text{eff}$ for the 
stochastic unraveling without approximation.

Since $H_\text{eff}$ is non-Hermitian, $\exp(-iH_\text{eff}t)\ket{\psi(0)}$ 
does not conserve the state's norm. As a consequence, for a given $\varepsilon$ 
drawn at random from a uniform distribution ${] 0,1]}$, there is a time, where 
the condition ${\left\lVert \psi(t) \rangle \right\rVert^2 \geq \varepsilon}$ 
is first violated. At this time, a jump with one of the Lindblad operators 
occurs, ${\ket{\psi(t)} \to \ket{\psi'(t)} =  L_j | \psi(t) \rangle/\left\lVert 
L_j | \psi(t) \rangle  \right\rVert}$, where the specific jump is chosen with 
probability ${p_j = \alpha_j \left\lVert L_j | \psi(t) \rangle 
\right\rVert^2/\sum_{j'} \alpha_{j'} \left\lVert L_{j'} |  \psi(t) \rangle 
\right\rVert^2}$. Thereafter, the next deterministic evolution with respect to 
$H_\text{eff}$ takes place. This sequence of stochastic jumps and deterministic 
evolutions leads to a particular trajectory. By averaging over trajectories, 
Eq.~\eqref{Eq::Lind1} can be approximated and expectation values follow as
\begin{equation}\label{Eq::ExpDeter}
\langle S_r^z(t) \rangle \approx \frac{1}{\text{T}_\text{max}} 
\sum_{\text{T}=1}^{\text{T}_\text{max}} \frac{\langle \psi_\text{T}(t) | S_r^z 
| \psi_\text{T}(t) \rangle}{\left\lVert |\psi_\text{T}(t) \rangle\right\rVert^2} 
\, , 
\end{equation}
where the subscript in $\ket{\psi_T(t)}$ labels a random sequence of jumps and 
deterministic evolutions.

\begin{figure}[tb]
\centering
\includegraphics[width=0.9\columnwidth]{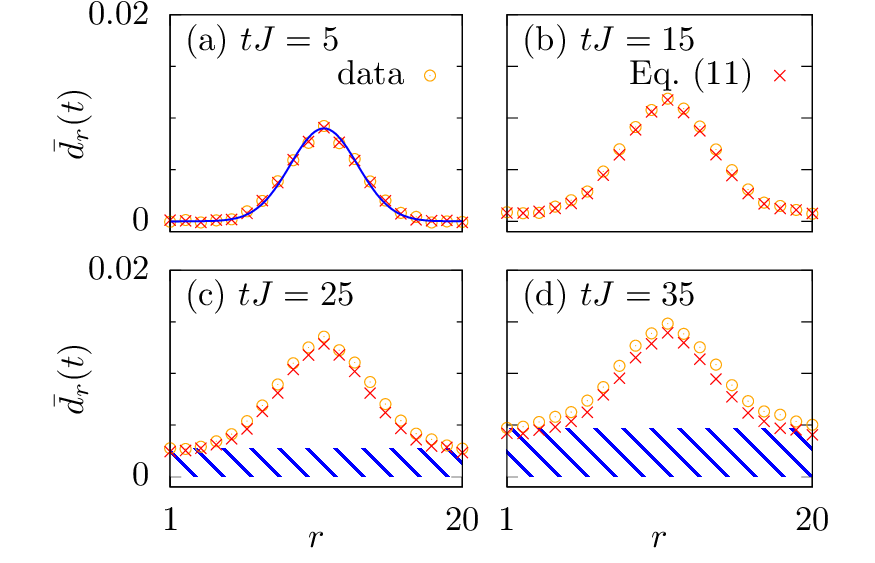}
\caption{Analogous setup as in Fig.~\ref{fig:peak}(b), but now for the full 
site dependence $\bar{d}_r(t)$ at various fixed times (a)-(d), which all lie in 
the middle of two jumps. Numerical data (circles) are in convincing agreement 
with the prediction in Eq.~(\ref{eq:superposition}) (crosses). A Gaussian is 
also indicated in (a) for comparison. The striped area indicates the 
equilibrium background of the already induced magnetization.}
\label{fig:profiles}
\end{figure}

\section{Dynamical typicality}
In a nutshell, quantum typicality asserts 
that a random pure quantum state can faithfully reproduce properties of the 
full statistical ensemble \cite{Gemmer2004, Goldstein2006, Popescu2006, 
Reimann2007, Bartsch2009}. For instance, a homogeneous magnetization 
distribution at $t = 0$, cf.\ Fig.~\ref{fig:variance}(b), can be readily
realized by a Haar-random initial state, $| \psi(0) \rangle = \sum_j (a_j + i 
b_j) \, | \phi_j \rangle$, where the coefficients $a_j$ and $b_j$ in some basis 
$| \phi_j \rangle$ are drawn at random from a Gaussian distribution with zero 
mean, and $\ket{\psi(0)}$ mimics the maximally mixed state $\rho \propto 1$ 
\cite{Gemmer2004, Goldstein2006, Popescu2006, Reimann2007, Bartsch2009}.
It is further instructive to consider, for the moment, an artificial scenario
with a quantum jump immediately at $t = 0$, i.e., $|\psi(0)\rangle \to
|\psi'\rangle \propto L_1\ket{\psi(0)}$. This results in a random superposition
over a subset of pure states with a spin-up at $r_0=N/2$, which mimics
${\rho \propto 1 + S_{r_0}^z}$. Then, the deterministic evolution $d_r(t)$ at 
weak driving, cf.\ Eq.~(\ref{Eq::TEVO}), 
\begin{equation}\label{Eq::DEVo}
d_r(t) \! \equiv \! \frac{\langle \psi'(t) | S_r^z | \psi'(t) 
\rangle}{\left\lVert| 
\psi'(t) \rangle \right\rVert^2} \! \approx \! \langle \psi' | e^{i H t} S_r^z  
e^{-i H t} | \psi' \rangle \, ,
\end{equation}
can be rewritten as ${d_r(t)/2 \approx \langle S_r^z(t) S_{r_0}^z(0)
\rangle}$ via typicality,  with ${S_r^z(t) = e^{i H t} S_r^z e^{-i H t}}$ and
${\langle \bullet \rangle = \text{tr}[\bullet]/2^N}$ denoting the 
infinite-temperature average \cite{Steinigeweg2017a}. Thus, the dynamics of
expectation values $d_r(t)$ during the deterministic process are generated by 
equilibrium correlation functions of the closed system $H$. We numerically 
demonstrate the validity of this finding in a test setting, where we consider
for simplicity only the single jump operator $L_1$ and artificially fix the 
jump times to ${\tau_k = k \delta \tau}$ with ${\delta \tau J = 10}$. As shown 
in Fig.~\ref{fig:peak}(a), $\langle S_r^z(t) S_{r_0}^z(0) 
\rangle$ indeed reproduces the deterministic dynamics after the first and 
before the next jump, $0 < t < 10$. Furthermore, Fig.~\ref{fig:peak}(a) already 
highlights that we can actually predict open-system trajectories even with many 
jumps, which is a main result of this work. As explained in the following, such 
a description of trajectories with multiple jumps is achieved by superimposing 
closed-system correlation functions $\langle S_r^z(t) S_{r_0}^z(0) \rangle$ 
appropriately. We should stress that the accuracy of the 
typicality approximation used so far increases exponentially with $N$ 
\cite{Steinigeweg2017a}.

\section{Connecting LRT and quantum trajectories}
To proceed, we now take into 
account also the jump operator $L_2$, but still use jump times $\tau_k = 
k\delta\tau$ for illustration. Averaging over trajectories weighted according 
to the jump probabilities of $L_1$ and $L_2$ with their different prefactors 
${\gamma (1+\mu)}$ and ${\gamma (1-\mu)}$, cf.\ text above 
Eq.~\eqref{Eq::ExpDeter}, one finds ${\bar{d}_r(t)/2 = \mu \langle S_r^z(t) 
S_{r_0}^z(0) \rangle}$ for the initial time evolution after the first 
jump, $0<t<10$, see Fig.~\ref{fig:peak}(b). While this idealized prediction 
cannot hold exactly at later stages of the trajectory, one can make further 
progress by assuming a sufficiently small value of $\gamma$. Then, within the 
deterministic evolution, the system has enough time to equilibrate and 
expectation values approach ${\bar{d}_r(t)/2 \to \mu \langle S_{r_0}^z(0)^2 
\rangle/N}$ [cf.\ Fig.~\ref{fig:peak}(b)], which 
approaches zero for large $N$ and thus becomes close to the local 
magnetizations before the initial jump at $t=0$. Eventually, another jump must 
occur at some time $\tau$ and, given the above equilibration, a reasonable 
expectation for the subsequent deterministic evolution is ${\bar{d}_r(t)/2 = 
\mu \langle S_{r_0}^z(0)^2 \rangle/N +  (\mu - \mu/N) \langle S_r^z(t - \tau) 
S_{r_0}^z(0) \rangle}$. Reiterating this procedure, we end up with a prediction 
for the entire trajectory with jump times $\tau_k$,
\begin{equation}
\bar{d}_r(t)/2 = \mu \sum_k A_k \, \Theta(t - \tau_k)
\, \langle 
S_r^z(t - \tau_k) S_{r_0}^z(0) \rangle \, , 
\label{eq:superposition}
\end{equation}
where $\Theta$ is the Heavyside function. The amplitudes $A_k$ read ${A_k/2 =
1/2 - \bar{d}_{r_0}(\tau_k - 0^+)/\mu}$ and measure the remaining deviation 
from the long-time equilibrium value, where we implicitly assumed full 
equilibration towards zero, via the balance ${\left\lVert L_1 | \psi(t) \rangle 
\right\rVert^2 = \left\lVert L_2 | \psi(t) \rangle \right\rVert^2}$. Equation 
(\ref{eq:superposition}) is the central result of this Letter. It predicts that 
the the open-system dynamics can be described by superimposing closed-system 
correlation functions at different times. Taking into account also an 
imbalance, i.e., ${\left\lVert L_1 | \psi(t)\rangle \right\rVert^2 \neq 
\left\lVert L_2 | \psi(t) \rangle \right\rVert^2}$, the $A_k$ can be further 
refined (see Appendix~\ref{app:amplitudes} for details),
\begin{equation}
\! \frac{A_k}{2} \! = \! 
\frac{a_k - \bar{d}_{r_0}(\tau_k - 0^+)}{\mu} \, , \,\,\, 
a_k \! = \! 
\frac{\mu - 2 \bar{d}_{r_0}(\tau_k - 0^+)}{2 - 4 \mu \, 
\bar{d}_{r_0}(\tau_k - 0^+)} \, , 
\label{eq:amplitudes}
\end{equation}
with $A_k \to 1$ if $\bar{d}_{r_0}(\tau_k - 0^+) \to 0$. In our numerics, we 
find Eqs.~(\ref{eq:superposition}) and (\ref{eq:amplitudes}) to be well 
fulfilled even if full equilibration is not reached, see Fig.~\ref{fig:peak}(b).
Importantly, Eq.~(\ref{eq:superposition}) not only applies at the bath site
$r_0=N/2$, but actually describes the full site dependence $\bar{d}_r(t)$ 
accurately, see Fig.~\ref{fig:profiles}, albeit with slight deviations at later 
times.

\begin{figure}[tb]
\centering
\includegraphics[width=0.9\columnwidth]{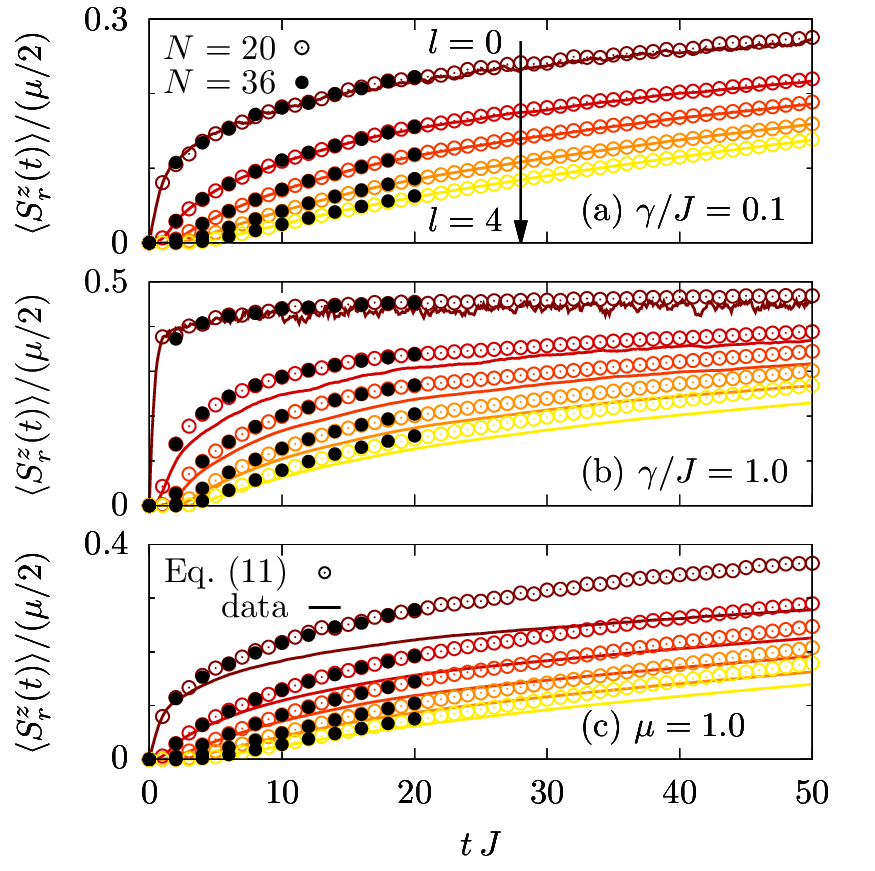}
\caption{Magnetization dynamics $\langle S_r^z(t) \rangle$ at different sites 
$r=r_0+l$ (curves), as generated by the full stochastic unraveling procedure 
(averaged over $10^5$ or more trajectories) for ${\Delta = 1.5}$ and ${N 
= 20}$. This procedure is performed for the full $H_\text{eff}$ without any 
approximation. (a) Small ${\gamma / J = 0.1}$ and (b) strong ${\gamma / J = 
1.0}$, both for weak ${\mu = 0.1}$. (c) Strong ${\mu = 1.0}$ and small ${\gamma 
/ J = 0.1}$. In all cases, we compare to the prediction 
(\ref{eq:superposition}) for ${N = 20}$ and ${N = 36}$ (circles).}
\label{fig:comparison}
\end{figure}

\section{From weak to strong driving}
While we have chosen artificial $\tau_k$ 
in Figs.~\ref{fig:peak} and \ref{fig:profiles} for illustrative reasons, we now 
turn to the actual solution of the Lindblad equation. Our analytical prediction 
for $\langle S_r^z(t)\rangle$ follows from averaging 
Eq.~\eqref{eq:superposition} over trajectories with random jump times 
(${\tau_1, \tau_2, \ldots\,}$), i.e., ${\langle S_r^z(t) \rangle \approx 
(1/\text{T}_\text{max}) \sum_{\text{T}} \bar{d}_{r, \text{T}}(t)}$. 
Specifically, given the exponential damping in Eq.~(\ref{Eq::TEVO}) for ${\mu 
\ll 1}$, the $\tau_k$ are given by ${\tau_{k+1} = \tau_k - \ln 
\varepsilon/\gamma}$, where a new $\varepsilon$ is drawn at random from 
${]0,1]}$ after each jump. Hence, if the correlation function ${\langle 
S_r^z(t) S_{r_0}^z(0) \rangle}$ is known, it is rather straightforward to 
construct the prediction (\ref{eq:superposition}) and the average numerically. 
Crucially, the computational costs of this procedure are significantly lower 
compared to the full stochastic unraveling such that we are able to generate 
dynamics for system sizes ${N = 36}$, see Fig.~\ref{fig:comparison} and 
Appendix~\ref{app:gamma}, which is approximately the 
maximum size reachable for typicality-based calculations of correlation 
functions. Note, however, that even larger system sizes might be reached when 
calculating correlation functions from matrix-product state techniques.

In Fig.~\ref{fig:comparison}(a)-(c), we summarize our numerical results 
for $\langle S_r^z(t)\rangle$, where we consider (i) weak driving ${\mu = 0.1}$ 
and weak coupling ${\gamma/J = 0.1}$, (ii) strong coupling ${\gamma/J = 1}$, 
and (iii) strong driving ${\mu = 1}$. We compare our prediction to the 
numerically exact stochastic unraveling which is performed for the full 
$H_\text{eff}$ and a homogeneous initial state. We find that the prediction
and the exact dynamics agree perfectly for (i), while the agreement becomes 
worse for (ii) and (iii), as expected. The convincing agreement in
Fig.~\ref{fig:comparison}(a) confirms our initial observation that the 
transport behavior of the closed system carries over to the open system (cf.\ 
Fig.~\ref{fig:variance}). Specifically, superpositions of correlation 
functions with diffusive (superdiffusive) scaling at ${\Delta = 1.5}$ (${\Delta 
= 1}$) according to Eq.~(\ref{eq:superposition}) yield a dynamics with the same 
scaling, see also Appendix~\ref{app:diffusioncoeff}.

\section{Conclusion}
In summary, we have studied nonequilibrium dynamics and 
transport in spin chains with a local coupling to a Lindblad bath. For weak  
driving, we have unveiled that the open-system dynamics can be constructed on 
the basis of closed-system correlation functions, which establishes a
connection between LRT and the Lindblad setting. For this specific setting, 
from a conceptual point of view, our results confirm the common assumption that 
closed-system and open-system approaches to transport should agree if the 
relevant parameters are chosen appropriately. From a practical point of view, 
our framework sheds new light on the efficient stochastic unravelings of 
Lindblad equations for large system sizes and long time scales. While we have 
chosen the XXZ chain as a timely example, our framework can be applied also to 
other spin or Hubbard models.

Promising directions of future research are, e.g., the generalization of our 
results to boundary-driven situations with a bath at each end of the system,
which seems to be feasible \cite{Heitmann2023}. Another interesting avenue is to 
study the role of integrability in more detail. In particular, our finding of 
persisting superdiffusive transport even in the presence of a system-bath 
coupling appears related to recent works that explored effect of weak 
integrability-breaking perturbations \cite{DeNardis2021, Wei2022}.

\section*{Acknowledgments} 
This work has been funded by the Deutsche Forschungsgemeinschaft (DFG), under 
Grants No.~397107022 (GE 1657/3-2) and  No.~397067869 (STE 2243/3-2), within 
the DFG Research Unit FOR 2692, under Grant No.~355031190. We acknowledge 
computing time at the HPC3 at University Osnabr\"uck, which has been funded by 
the DFG under Grant No.~456666331. Jonas Richter acknowledges funding from the 
European Union's Horizon Europe research and innovation programme
under the Marie Sk\l odowska-Curie grant agreement
No. 101060162, and the Packard Foundation through a Packard
Fellowship in Science and Engineering.


\begin{appendix}

  \section{Amplitudes}\label{app:amplitudes}

  One possibility to derive the amplitudes in Eq.~\eqref{eq:amplitudes} 
  is based 
  on typicality arguments. To this end, consider a maximally random pure state 
  ${| \psi(\tau_j - 0^+) \rangle}$ under the constraint
  \begin{equation}
    d_{r_0}(\tau_j - 0^+) = x \, ,
  \end{equation}
  before a jump occurs at time $\tau_j$. Then, we have
  \begin{equation}
  y_1 = \left\lVert L_1 | \psi(\tau_j - 0^+) \rangle \right\rVert^2 = \frac{1}{2} - x
  \end{equation}
  and
  \begin{equation}
  y_2 = \left\lVert L_2 | \psi(\tau_j - 0^+) \rangle \right\rVert^2 = x + \frac{1}{2}
  \end{equation}
  with ${y_1 + y_2 = 1}$. The corresponding jump probabilities read
  \begin{equation}
  p_1 = \frac{(1+\mu) y_1}{(1+\mu) y_1 + (1-\mu) y_2}
  \end{equation}
  and
  \begin{equation}
  p_2 = \frac{(1-\mu) y_2}{(1-\mu) y_2 + (1+\mu) y_1}
  \end{equation}
  with ${p_1 + p_2 = 1}$ again. Consequently, a straightforward calculation yields
  \begin{equation}
  \frac{p_1}{2} - \frac{p_2}{2} = \frac{\mu - 2 x}{2 - 4 \mu x} \, ,
  \end{equation}
  i.e., the expression in Eq.~\eqref{eq:amplitudes}.
  
  \section{Dependence on $\gamma$ and $N$}
  \label{app:gamma}

  \begin{figure}[tb]
    \centering
    \includegraphics[width=0.9\columnwidth]{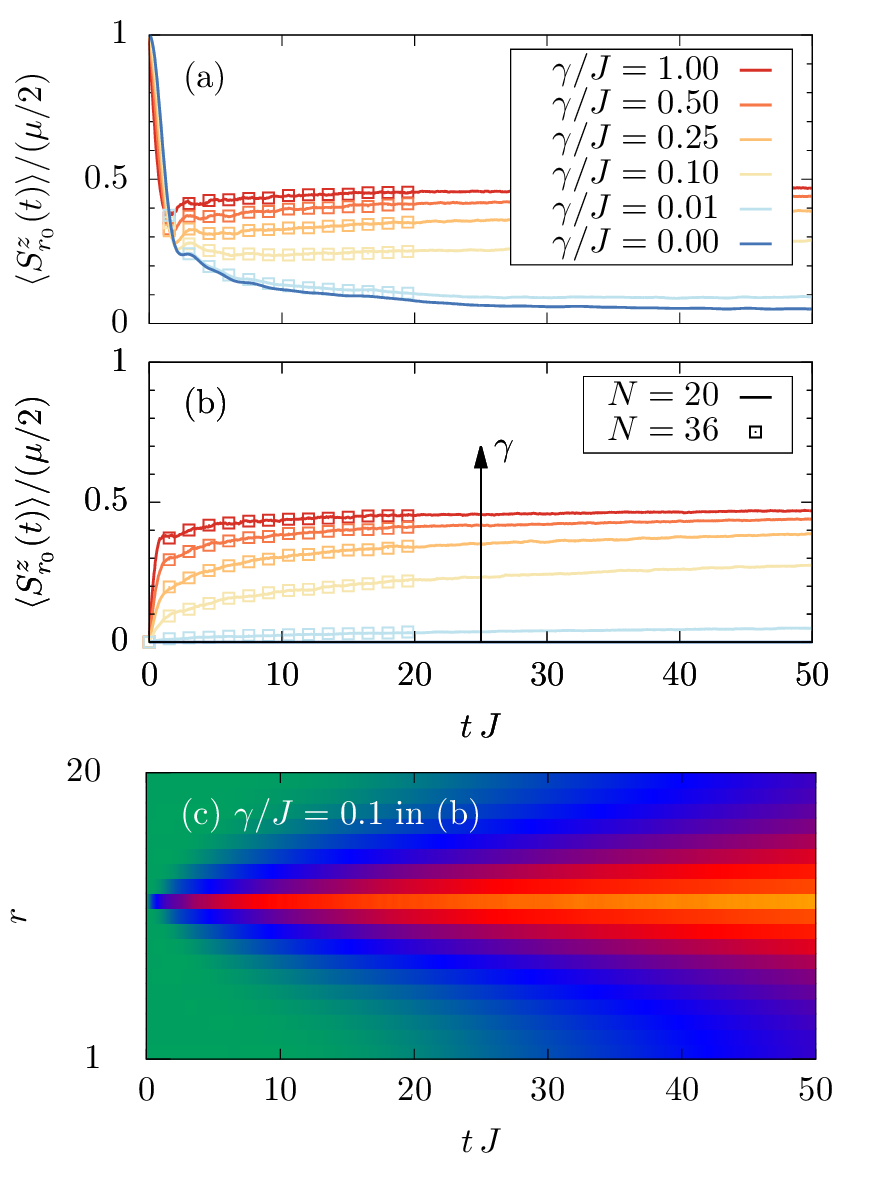}
    \caption{Time evolution of the magnetization 
    ${\langle S_{r_0}^z(t) \rangle}$ 
    at the position $r_0=N/2$ 
    of the local Lindblad operators, as given for 
    weak driving ${\mu = 0.1}$ by Eq.~\eqref{eq:superposition} with amplitudes
    according to Eq.~\eqref{eq:amplitudes}. 
    Curves for various values of the bath 
    coupling $\gamma$ are obtained from the average over 10,000 different 
    trajectories. The other model parameters are the same as in 
    Figs.~\ref{fig:peak} and \ref{fig:profiles}. 
    A bath coupling ${\gamma/J = 0.1}$ is 
    comparable to the jump times in Fig.~\ref{fig:peak}. (a) and (b) correspond to 
    an initial state with and without local magnetization, respectively. In each 
    case, data for ${N = 36}$ sites is also depicted. (c) Full site dependence for 
    ${\gamma/J = 0.1}$ in (b).}
    \label{fig:prediction}
    \end{figure}
  
  Since we have mostly discussed the case of a small bath coupling ${\gamma / J = 
  0.1}$, we depict in Fig.~\ref{fig:prediction} the prediction according to 
  Eq.~\eqref{eq:superposition} for various values of $\gamma$. We do so for the 
  magnetization ${\langle S_{r_0}^z(t) \rangle}$ 
  at the position $r_0=N/2$ of the local 
  Lindblad operators 
  and random initial states ${| \psi(0) \rangle}$ with and without local 
  magnetization. Moreover, to demonstrate that this prediction does not depend 
  on system size, we also show the corresponding prediction for ${N = 36}$ sites.
  
  \begin{figure}[t]
  \centering
  \includegraphics[width=0.9\columnwidth]{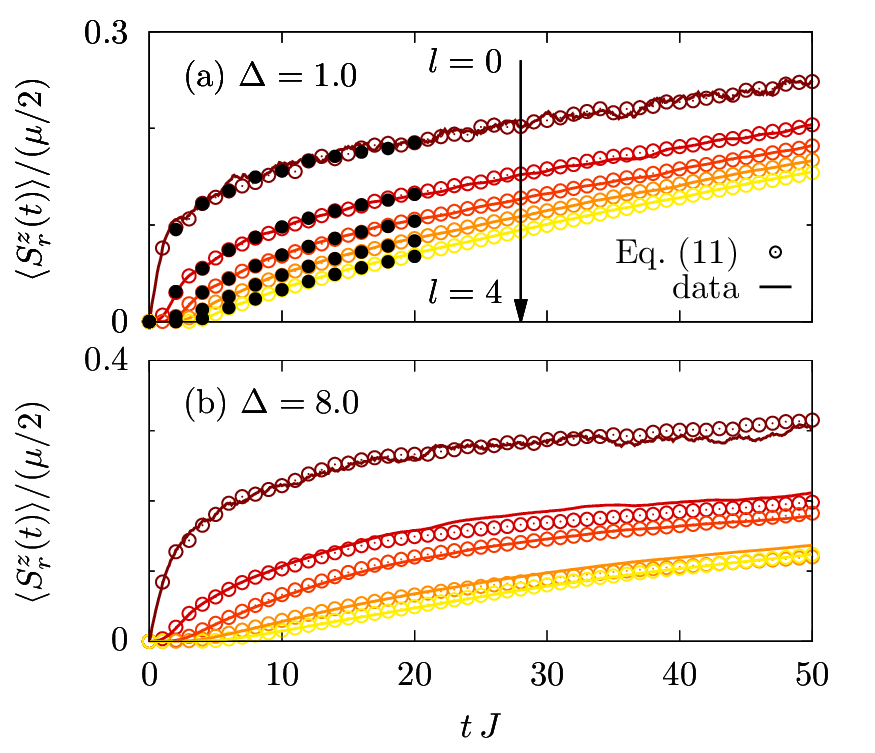}
  \caption{Dynamics of the magnetization $\langle S_r^z(t) 
  \rangle$ at various sites ${r=r_0+l}$ 
  for anisotropies (a) ${\Delta = 1.0}$ and (b) 
  ${\Delta = 8.0}$, as generated by the full stochastic unraveling procedure and as 
  predicted by Eq.~\eqref{eq:superposition}. Remaining model parameters: Small 
  coupling ${\gamma/J = 0.1}$, weak driving ${\mu = 0.1}$, 
  and system size ${N = 20}$.}
  \label{fig:anisotropies}
  \end{figure}
  
  \begin{figure}[b]
  \centering
  \includegraphics[width=0.9\columnwidth]{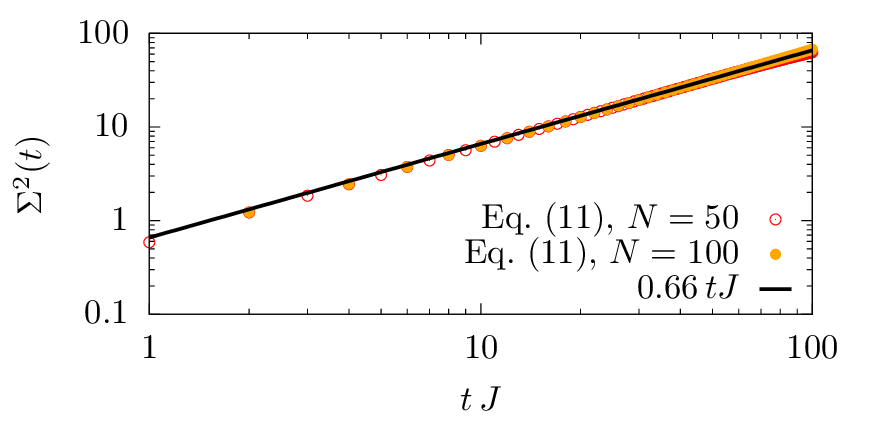}
  \caption{Time-dependent spatial variance $\Sigma^2(t)$, as 
  predicted by Eq.~\eqref{eq:superposition} for ${D_\text{closed} /J = 0.6}$.}
  \label{fig:perfect}
  \end{figure}
  
  \section{Other anisotropies}
  \label{app:delta}
  
  In Fig.~\ref{fig:comparison}, 
  we have provided a detailed comparison of the 
  dynamics for anisotropy ${\Delta = 1.5}$, as generated by the full stochastic 
  unraveling procedure and as predicted by Eq.~\eqref{eq:superposition}. To 
  demonstrate that an agreement of similar quality can be obtained for other 
  anisotropies as well, we show in Fig.~\ref{fig:anisotropies} a comparison for 
  ${\Delta = 1.0}$ and ${\Delta = 8.0}$, in both cases for small coupling 
  ${\gamma / J = 0.1}$ and weak driving ${\mu = 0.1}$.
  
  \section{Diffusion coefficient}
  \label{app:diffusioncoeff}
  
  Let us, for simplicity, estimate the expansion velocity of the open 
  system by
  \begin{equation}
  \frac{v_\text{open}}{D_\text{closed}} = \frac{D_\text{closed} (t - 
  \bar{\tau})}{D_\text{closed}\, t}
  \end{equation}
  with the average injection time
  \begin{equation}
  \bar{\tau} = - \frac{1}{\gamma} \int_{0^+}^1  \text{d} \varepsilon \, \ln 
  \varepsilon \, ,
  \end{equation}
  which is ${\bar{\tau} J \approx 10 }$ for ${\gamma / J = 0.1}$. By taking into 
  account ${D_\text{closed} / J \approx 0.6}$ for ${\Delta = 1.5}$, one would expect 
  at ${t = 2 \bar{\tau}}$ the expansion velocity
  \begin{equation}
  \frac{v_\text{open}}{J} \approx \frac{0.6}{2} = 0.3 \, .
  \end{equation}
  Thus, a reasonable expectation is
  \begin{equation}
  \Sigma^2(t) = 2 \, v_\text{open} t \approx 0.6 \, t J \, .
  \end{equation}
  And indeed, this number is chosen as the prefactor of the power law in 
  Fig.~\ref{fig:variance}.
  
  \begin{figure}[b]
  \centering
  \includegraphics[width=0.9\columnwidth]{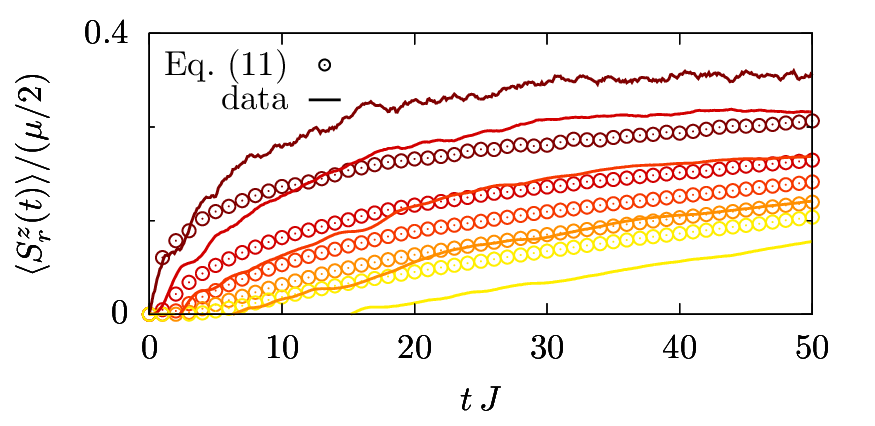}
  \caption{Analogous comparison as the one in 
  Fig.~\ref{fig:comparison}(a), 
  but now the initial pure state ${| \psi(0) \rangle}$ is 
  not drawn at random.}
  \label{fig:untypical}
  \end{figure}
  
  An alternative and kind of better way to estimate the expansion velocity in 
  the open system is provided by Eq.~\eqref{eq:superposition} and the assumption 
  of perfectly diffusive behavior in the closed system (with a zero mean free 
  path). Then, the equilibrium correlation functions take on the simple form
  \begin{equation}
  \langle S_r^z(t) S_{r_0}^z(0) \rangle  = 
  \frac{1}{4} \, \exp (-2 D_\text{closed} t) 
  \, {\cal I}_r(2 D_\text{closed} t) \, ,
  \end{equation}
  where ${\cal I}_r(t)$ is the modified Bessel function of the first kind
  and of the order $r$. By the use of this assumption, the calculation of the 
  time-dependent variance $\Sigma^2(t)$ in the open system can be done 
  numerically. As depicted in Fig.~\ref{fig:perfect} for 
  ${D_\text{closed} /J = 0.6}$, one finds
  \begin{equation}
  \Sigma^2(t) \approx 0.66 \, t J
  \end{equation}
  over a wide range of time, which is consistent with the simple argument above. 
  Note that the calculation can be easily carried out for ${N = 100}$ of lattice 
  sites.
  
  \section{Other initial states}
  \label{app:untypical}
  
  The derivation of the prediction in Eq.~\eqref{eq:superposition}
  has relied on an initial pure state ${| \psi(0) \rangle}$, 
  which is fully random and corresponds to an equilibrium density matrix at 
  formally infinite temperature. In Fig.~\ref{fig:untypical}, we demonstrate that 
  this prediction does not apply to other initial states. To this end, we choose 
  the specific initial pure state
  \begin{equation}
  | \psi(0) \rangle \propto (| \! \uparrow \rangle +  | \! \downarrow \rangle) 
  \otimes \ldots \otimes (| \! \uparrow \rangle +  | \! \downarrow \rangle) \, ,
  \end{equation}
  which is known to be untypical.

\end{appendix}

\bibliographystyle{apsrev4-1_titles}
\bibliography{paper.bib}

\end{document}